\documentclass[%
 reprint,
 amsmath,amssymb,
 aps,
prb,
]{revtex4-2}

\usepackage[version=3]{mhchem} 
\usepackage{graphicx}
\usepackage{bm}
\usepackage{amsmath}
\usepackage{amssymb}
\usepackage{color}

\begin{document}

\author{Rafaela F. S. Penacchio}
\affiliation{Institute of Physics, University of S{\~{a}}o Paulo, 05508-090 S{\~{a}}o Paulo, SP, Brazil}

\author{Celso I. Fornari}
\affiliation{Experimentelle Physik VII and W{\"u}rzburg Dresden Cluster of Excellence ct.qmat, Fakult{\"a}t f{\"u}r Physik und Astronomie, Universit{\"a}t W{\"u}rzburg, Am Hubland, D-97074 W{\"u}rzburg, Germany}

\author{Yor{\'i} G. Camillo}
\affiliation{Institute of Physics, University of S{\~{a}}o Paulo, 05508-090 S{\~{a}}o Paulo, SP, Brazil}

\author{Philipp Kagerer}
\affiliation{Experimentelle Physik VII and W{\"u}rzburg Dresden Cluster of Excellence ct.qmat, Fakult{\"a}t f{\"u}r Physik und Astronomie, Universit{\"a}t W{\"u}rzburg, Am Hubland, D-97074 W{\"u}rzburg, Germany}

\author{Sebastian Buchberger}
\affiliation{Experimentelle Physik VII and W{\"u}rzburg Dresden Cluster of Excellence ct.qmat, Fakult{\"a}t f{\"u}r Physik und Astronomie, Universit{\"a}t W{\"u}rzburg, Am Hubland, D-97074 W{\"u}rzburg, Germany}

\author{Martin Kamp}
\affiliation{Physikalisches Institut and R{\"{o}}ntgen-Center for Complex Material Systems (RCCM), Fakult{\"{a}}tf{\"{u}}r Physik und Astronomie, Universit{\"{a}}tW{\"{u}}rzburg, W{\"{u}}rzburg D-97074, Germany}

\author{Hendrik Bentmann}
\affiliation{Experimentelle Physik VII and W{\"u}rzburg Dresden Cluster of Excellence ct.qmat, Fakult{\"a}t f{\"u}r Physik und Astronomie, Universit{\"a}t W{\"u}rzburg, Am Hubland, D-97074 W{\"u}rzburg, Germany}

\author{Friedrich Reinert}
\affiliation{Experimentelle Physik VII and W{\"u}rzburg Dresden Cluster of Excellence ct.qmat, Fakult{\"a}t f{\"u}r Physik und Astronomie, Universit{\"a}t W{\"u}rzburg, Am Hubland, D-97074 W{\"u}rzburg, Germany}

\author{S{\'{e}}rgio L. Morelh{\~{a}}o}
\affiliation{Institute of Physics, University of S{\~{a}}o Paulo, 05508-090 S{\~{a}}o Paulo, SP, Brazil}
\email{morelhao@if.usp.br}

\title{X-ray diffraction tools for structural modeling of epitaxic films of an intrinsic antiferromagnetic topological insulator}


\begin{abstract}
Synthesis of new materials demands structural analysis tools suited to the particularities of each system. Van der Waals (vdW) materials are fundamental in emerging technologies of spintronics and quantum information processing, in particular topological insulators and, more recently, materials that allow the phenomenological exploration of the combination of non-trivial electronic band topology and magnetism. Weak vdW forces between atomic layers give rise to composition fluctuations and structural disorder that are difficult to control even in a typical binary topological insulators such as \ce{Bi2Te3}. The addition of a third element as in \ce{MnBi2Te4} makes the epitaxy of these materials even more chaotic. In this work, statistical model structures of thin films on single crystal substrates are described. It allows the simulation of X-ray diffraction in disordered heterostructures, a necessary step towards controlling the epitaxial growth of these materials. On top of this, the diffraction simulation method described here can be readily applied as a general tool in the field of design new materials based on stacking of vdW bonded layers of distint elements.
\end{abstract}

\maketitle
\section{Introduction}\label{intro}

Two-dimensional (2D) van der Waals (vdW) materials have experienced an explosive growth after graphene, and new families of 2D systems and block-layered bulk materials have been discovered \cite{mo19b,wm21,lh21}. The possibility of tuning their electronic properties via structural parameters make the layered vdW materials attractive from both fundamental and device engineering points of view. This field has become particularly interesting after the experimental discovery of three-dimensional (3D) topological insulators (TIs), having as a prototypical the bismuth chalcogenide compounds \cite{yc09,hz09,yx09,dh09}. To control the chemical potential of these compounds without using extrinsic doping, growth methods and properties of thin films have been investigated \cite{yl10,gw11,kh14,yg15}. However, the weakness of vdW interlayer forces lead in general to systems undergoing drastic changes as a function of subtle variation in growth conditions. Finding controllable fabrication processes of such systems has proven challenging \cite{sm19,cf20a}. On top of this, the recently discovered intrinsic magnetic topological insulator \ce{MnBi2Te4} has added a new chapter to the phenomenological exploration of combining non-trivial electronic band topology and magnetism. Contrary to other attempts of breaking time-reversal symmetry by diluted doping of transition metals or rare-earth elements on 3D TIs \cite{cc13,cc15,cf20,af10,sh15}, this material carries in its unit cell ordered layers of Mn atoms, providing a ferromagnetic ordering in the plane and a broad range of out-of-plane configurations depending on the stacking sequence \cite{mo19a,jl19}. This compound is part of the (\ce{MnBi2Te4})$_n$(\ce{Bi2Te3})$_m$ homologous series, similar to the (\ce{Bi2Te3})$_n$(\ce{Bi2})$_m$ series \cite{jb07,cf16a,hs14}. The series is composed by stacking two fundamental building blocks and spans from \ce{Bi2Te3} ($n = 0$), the archetypal of 3D TI without magnetic ordering, to the intrinsic antiferromagnetic \ce{MnBi2Te4} ($m = 0$), passing through an infinity of intermediary phases \cite{za19,pk20,er19}.

For the 3D non‑magnetic TI, \ce{Bi2Te3}, the unit cell is composed by stacking three quintuple layers (QLs). These QLs are fundamental building blocks always started and terminated in Te atoms as Te-Bi-Te-Bi-Te. The Te-Bi atoms are ionic bonded inside the QLs, while these blocks are coupled together along the [0001] direction due to weak vdW forces between the Te atoms in adjacent blocks. By inserting Mn in this structure, an extra MnTe double layer is formed inside the QL, leading to the existence of septuple layers (SLs) as Te-Bi-Te-Mn-Te-Bi-Te. The atomic Mn layer inside the SLs present a net out-of-plane magnetic moment, that is, the neighbor atoms are ferromagnetically coupled inside the layer. When stacked together, the SLs present anti-ferromagnetic (AFM). The intermediary phases are determined by the ratio of SLs and QLs, and present interesting AFM properties \cite{rv19}. Besides the above mentioned magnetic properties of this series, an even richer interplay between topology and magnetism is expected in the 2D regime when reducing the number of stacked layers \cite{mo19b}. Such properties make this intrinsic magnetic TI highly attractive. However, a great challenge rising in this field is to control and understand the growth mechanisms of this compound in order to suppress the formation of structural defects and prepare perfectly ordered layers to explore the 3D to 2D transition.

Simulations of X-ray scattering and diffraction are well-established procedures for structural analysis at nanometer and subnanometer length scales of layered materials, ranging from amorphous films to crystalline ones such as epitaxial layers on single-crystal substrates \cite{sm02b,sm17}. Higher are the ordering in stacking sequences of the atomic layers, the more pronounced are the diffracted intensities at higher angles allowing more refined structure models. X-ray theories are well comfortable at the limiting cases, either amorphous films or perfect periodic layer sequences, that is crystalline films. However, in developing new materials and processing technologies, layered materials with random layer sequences of large $d$-spacing can often be found. Combined with the very high dynamical range of advanced X-ray sources and instruments, this kind of material represent a challenging in theoretical approach for X-ray diffraction simulation \cite{sm11,aa06}. By using state-of-the-art X-ray diffraction simulation  applied to the (\ce{Bi2Te3})$_m$(\ce{MnBi2Te4})$_n$ homologous series, it is possible to study details of the stacking layers, providing information to further improvements on the growth of this material.

In this work, we describe how to adapt a general recursive equation for simulating X-ray dynamical diffraction in layered materials to the case of thin films of Mn$_x$Bi$_2$Te$_{3+x}$ (MBT) grown by molecular beam epitaxy on BaF$_2$ (111) substrates. The films are stacks of $n$ vdW bounded MnBi$_2$Te$_4$ septuple layers where the occurrence of $m$ Bi$_2$Te$_3$ quintuple layers leads to films of composition $x = n/(n+m)$. Interface quality, random stacking sequences, surface finishing, and evolution of defects during growth are accessible parameters by curve fitting with the recursive equation. The effectiveness of this approach is demonstrated for analysing X-ray diffraction in films with different compositions and disorder parameter, ranging from random to perfect periodic stacking sequences of atomic layers. 

\section{Structure Models}\label{sec:SM}

\begin{figure*}
    \centering
    \includegraphics[width=0.75\textwidth]{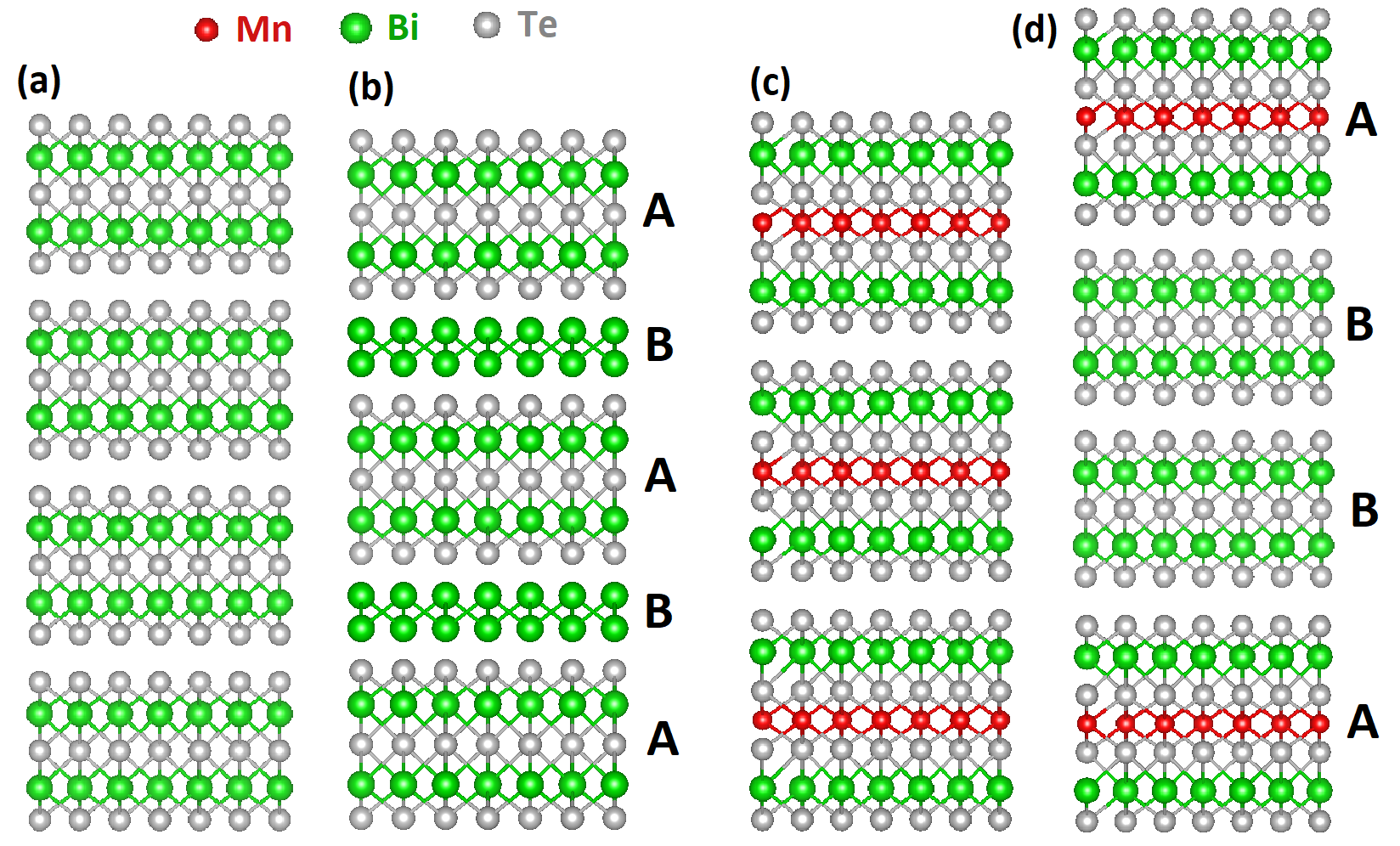}
    \caption{Building blocks in Bi$_2$Te$_{3-\delta}$ and MBT epitaxic films. (a) Pure Bi$_2$Te$_{3}$ phase with no Te deficit ($\delta=0$). (b) Bilayers of bismuth as in (Bi$_2$Te$_{3}$)$_n$(Bi$_2$)$_m$ $=$ $A_n B_m$ lead to films with Te defict $\delta=3m/(n+m)$; main and secondary building blocks: $A={\rm Te}\!:\!{\rm Bi}\!:\!{\rm Te}\!:\!{\rm Bi}\!:\!{\rm Te}$ and $B={\rm Bi}\!:\!{\rm Bi}$. (c) Single MnBi$_2$Te$_{4}$ phase. (d) Mixing of phases (MnBi$_2$Te$_4$)$_n$(Bi$_2$Te$_3$)$_m$ $=$ $A_n B_m$ lead to Mn$_x$Bi$_2$Te$_{3+x}$ films with composition $x=n/(n+m)$ or Mn deficit $\gamma = 1-x = m/(n+m)$; main and secondary building blocks: $A={\rm Te}\!:\!{\rm Bi}\!:\!{\rm Te}\!:\!{\rm Mn}\!:\!{\rm Te}\!:\!{\rm Bi}\!:\!{\rm Te}$ and $B={\rm Te}\!:\!{\rm Bi}\!:\!{\rm Te}\!:\!{\rm Bi}\!:\!{\rm Te}$.}
\label{fig:ABblocks}
\end{figure*}

A key point in vdW epitaxy is the weakness of interlayer forces. It favours, in principle, flexibilization of the lateral lattice matching requirements \cite{ag13,pv18} at the same time that makes challenging the control of film composition and other lattice defects \cite{jh17,gs18,mi19,sm19}. Modeling disordered heterostructures is a necessary step towards general procedures for structural analysis of materials based on vdW epitaxy. Composition fluctuation is related to the occurrence of distinct building blocks---sets of atomic monolayers sharing covalent bonds---randomly stacked along film thickness and bonded to each other by weak vdW forces. Figures~\ref{fig:ABblocks}a and \ref{fig:ABblocks}b show the building blocks that have been used for modeling Bi$_2$Te$_{3-\delta}$ films with deficit $\delta$ of tellurium \cite{sm17}, and in Figures~\ref{fig:ABblocks}d \ref{fig:ABblocks}c, the blocks that are used here for modeling the MBT films. 

Accounting for differences in the general appearance of the films with composition fluctuation can be accomplished by correctly choosing a probability function to control the stacking sequences of two building blocks; the main block labeled $A$ and the secondary one labeled $B$.
For instance, in one case both types of building blocks can occur adjacently to each, as in the sequence $A\!:\!A\!:\!B\!:\!B\!:\!A\!:\!A\!:\!A\!:\!B\!:\!A$, and in the other case it occurs to one type of block only, as in $A\!:\!A\!:\!B\!:\!A\!:\!A\!:\!A\!:\!B\!:\!A\!:\!A$. Even in cases where the synthesis is aimed at obtaining materials with only the main block $A$, it is necessary to know how to evidence and quantify, if possible, the occurrence of the other block, block $B$, that is responsible for composition fluctuation. In more refined models, changes in interlayer distances as a function of composition can be taken into account, as in the case of Bi$_2$Te$_{3-\delta}$ films \cite{hs14,cf16a}. However, in new materials as the MBT epitaxic films \cite{pk20}, further improvement in film quality will be necessary before resolving the variations in interlayer distances with composition. 

In structure models, a stack of $N$ adjacent $A$ blocks---without a $B$ block in the middle---occurs with probability  
\begin{equation}\label{eq:PofN}
  P(N) = \int_{N-1/2}^{N+1/2} p(z) {\rm d}z
\end{equation}
where $N$ stands for non-negative integers. Therefore, films made exclusively of $A$ blocks have $P(N) = 1$ for $N$ as the total number of blocks in the film structure along the growth direction. To account for films where the secondary $B$ blocks can be adjacent to $A$ blocks only, the probability function based on a standard log-normal function 
\begin{equation}\label{eq:Lz}
   p(z) = \frac{1}{z\, \sigma_L \sqrt{2\pi}}\,{\rm e}^{-\frac{1}{2}\left[\frac{{\rm ln}(z)-{\rm ln}(b)}{\sigma_L}\right]^2}\,,
\end{equation}
can describe well the stacking sequences from totally disordered to a perfectly periodic heterostructure. $b=N_0{\rm exp}(\sigma_L^2)$, $N_0$ is the mode, and $\sigma_L$ the standard deviation in log-scale. To assure that there will be no $B\!:\!B$ sequences in the films, that is $P(0)=0$, the small value of $p(z)$ in the range from $z=0$ to $1/2$ is accounted for in $P(1) = \int_{0}^{3/2} p(z) {\rm d}z$. The closest two $B$ blocks can be to each other is in $B\!:\!A\!:\!B$, which occurs with probability $P(1)$. 

\begin{figure*}
    \centering
    \includegraphics[width=\textwidth]{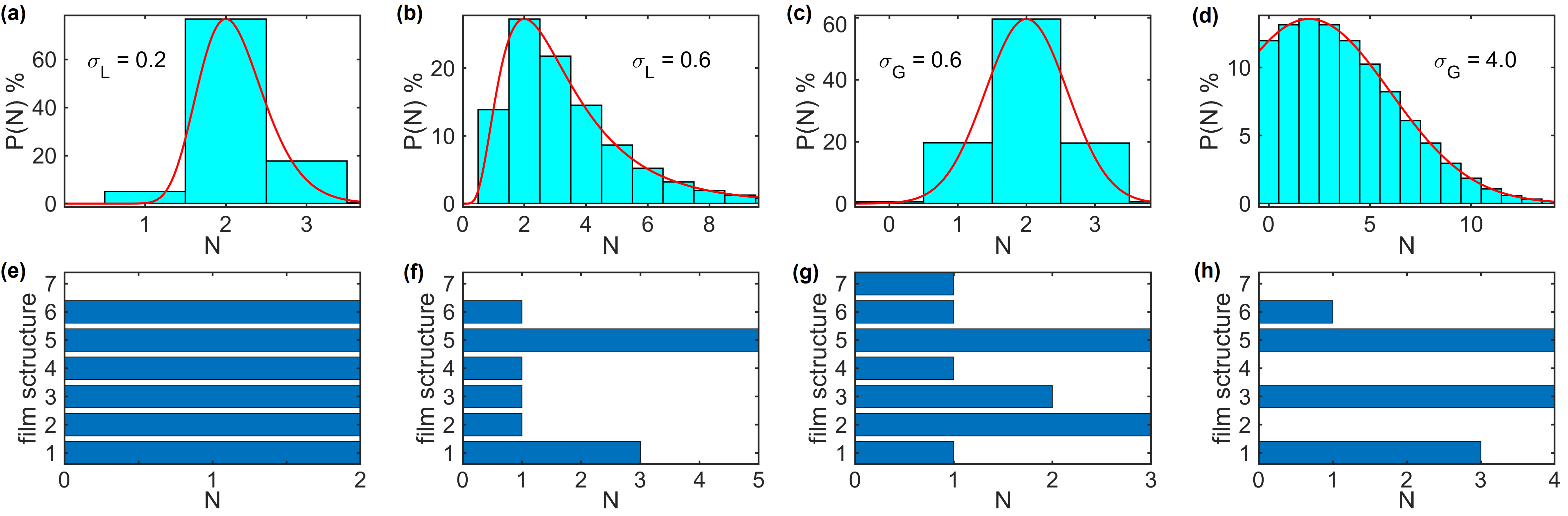}
    \caption{(a-d) Probability functions $P(N)$, eq.~(\ref{eq:PofN}), for stacking sequences of $N$ adjacent $A$ blocks interleaved by $B$ blocks.  (e-h) Graphic schemes of $A_{12}B_6$ film structures along thickness according to the probability showed above each scheme: stacks of $N$ adjacent $A$ blocks (blue bars) separated by $B$ block (vertical gap in between the blue bars). (h) Disordered structure with adjacent $B$ blocks (large vertical gap). $P(0)$ gives the probability for the absence of one $A$ block in between two $B$ blocks.}
\label{fig:distfunc}
\end{figure*}

Examples of periodic and disordered sequences are shown in Figure~\ref{fig:distfunc}. For a mode $N_0 = 2$ and a narrow deviation such as $\sigma_L=0.2$, the distribution of probability in Figure~\ref{fig:distfunc}a gives about 5\% for isolated $A$, 77\% for $A\!:\!A$, and 18\% for $A\!:\!A\!:\!A$. The chances for generating a highly periodic $A_n B_m$ heterostructure is given by $P(N_0)^{n/N_0}$, as the 12:6 ($n\!\!:\!\!m$) film in Figure~\ref{fig:distfunc}e with stacking sequence $A\!:\!A\!:\!B\!:\!A\!:\!A\!:\!B\!:\!A\!:\!A\!:\!B\!:\!A\!:\!A\!:\!B\!:\!A\!:\!A\!:\!B\!:\!A\!:\!A\!:\!B$ that occurs with probability $P(2)^{6} \simeq 20\%$ within an ensemble of structure models. Perfectly periodic heterostructures also require $N_0=n/m$ to be an integer number. On the other hand, when a broad probability function is used for structure models generation, such as the one with $\sigma_L=0.6$ in Figure~\ref{fig:distfunc}b, disordered heterostructures are highly probable, as in Figure~\ref{fig:distfunc}f with sequence $A\!:\!A\!:\!A\!:\!B\!:\!A\!:\!B\!:\!A\!:\!B\!:\!A\!:\!B\!:\!A\!:\!A\!:\!A\!:\!A\!:\!A\!:\!B\!:\!A\!:\!B$. Despite the degree of disorder, there is no $B\!:\!B$ sequences as the probability function was chosen to provide $P(0)=0$. This probability function with $p(z)$ in eq.~(\ref{eq:Lz}) has been used for modeling Bi$_2$Te$_{3-\delta}$ films, as detailed described elsewhere \cite{sm17}. 

In epitaxic films where both types of building blocks can form stacks of equal blocks, that is where $B$ blocks can appear adjacently to each other, the probability function
has to allow $P(0)>0$. It can be accomplish by using, for instance, a Gaussian-based function such as 
\begin{equation}\label{eq:Gz}
    p(z) = \frac{{\rm e}^{  -\frac{1}{2}\left(\frac{z-N_0}{\sigma_G}\right)^2 }}{\int \limits_{-1/2}^{\infty} {\rm e}^{  -\frac{1}{2}\left(\frac{z-N_0}{\sigma_G}\right)^2 } {\rm d}z}
\end{equation}
with mode $N_0$ and standard deviation $\sigma_G$. Narrow probability functions, as in Figure~\ref{fig:distfunc}c where $P(0)\approx0$, lead to more ordered distribution of the $B$ blocks, as in the sequence $A\!:\!B\!:\!A\!:\!A\!:\!A\!:\!B\!:\!A\!:\!A\!:\!B\!:\!A\!:\!B\!:\!A\!:\!A\!:\!A\!:\!B\!:\!A\!:\!B\!:\!A$ shown in Figure~\ref{fig:distfunc}g. However, broader probability functions,  as in Figure~\ref{fig:distfunc}d, increase the disorder of the $B$ blocks and also the probability of some of them to appear together as in the sequence $A\!:\!A\!:\!A\!:\!B\!:\!B\!:\!A\!:\!A\!:\!A\!:\!A\!:\!B\!:\!B\!:\!A\!:\!A\!:\!A\!:\!A\!:\!A\!:\!B\!:\!A\!:\!B$ graphically represented in Figure~\ref{fig:distfunc}h.

\begin{figure*}
    \centering
    \includegraphics[width=\textwidth]{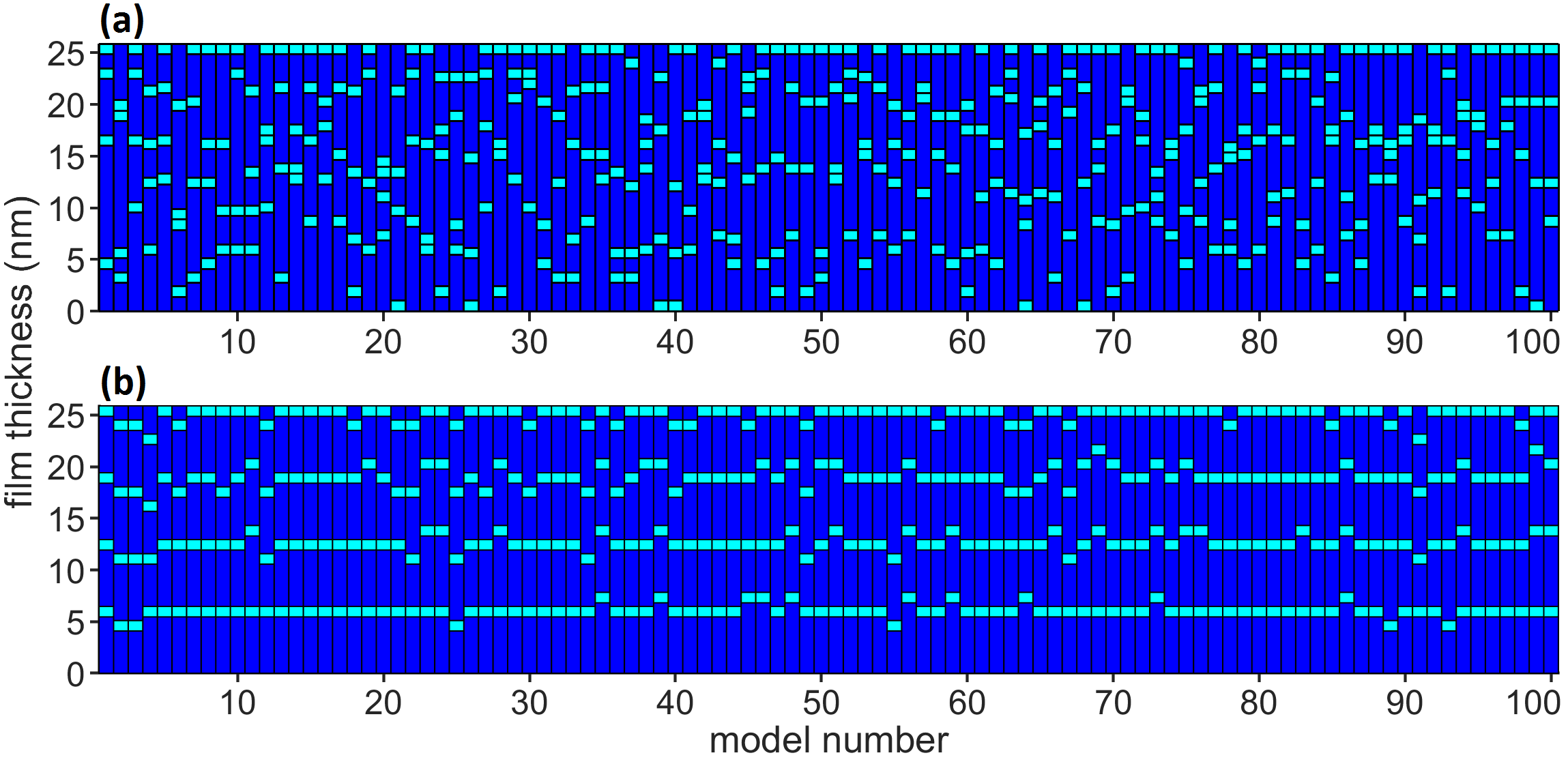}
    \caption{Ensembles of hundred structure models for $A_{16}B_4$ films where $A = {\rm MnBi}_2{\rm Te}_4$ (blue, 1.36\,nm) and $B = {\rm Bi}_2{\rm Te}_3$ (cyan, 1.02\,nm). (a) $\sigma_G = 4$ and (b) $\sigma_G = 0.4$ in eq.~(\ref{eq:Gz}).Total film thickness is 25.8\,nm for all models. }
\label{fig:filmstructure}
\end{figure*}

\section{X-ray Diffraction Simulation}
\label{sec:xrd}

For a given probability function and film composition, hundreds of structure models are generated with constant numbers of both $A$ and $B$ blocks, as exemplified in Figure~\ref{fig:filmstructure}. Such ensemble of models represents possible statistical fluctuation within film domains distributed over the sample area. X-ray reflectivity of each model is calculated by adding up reflection and transmission coefficients of the successive layers along film thickness. If $r_{X,Y}$ and $t_{X,Y}$ stand for reflection and transmission coefficients of generic $X$ and $Y$ layers, the coefficients of the combined $Y\!:\!X$ layer with $X$ on top of layer $Y$ are calculated according to \cite{sm17}
\begin{eqnarray}\label{eq:receqs}
  r_{YX} &=& r_X+r_Y\frac{t_X t_X\,e^{2 i \varphi}}{1 - \bar{r}_X r_Y\,e^{2 i \varphi}}\;, \nonumber \\
  \bar{r}_{YX} &=& \bar{r}_Y+\bar{r}_X\frac{t_Y t_Y\,e^{2 i \varphi}}{1 - \bar{r}_X r_Y\,e^{2 i \varphi}}\;,\quad{\rm and}  \\
  t_{YX} &=& \frac{t_X t_Y\,e^{i \varphi}}{1 - \bar{r}_X r_Y\,e^{2i \varphi}}\;.\nonumber
\end{eqnarray}
$\varphi=-\frac{1}{2}Qd$ is the phase delay every time the X-ray wave of wavelength $\lambda$ crosses the interlayer distance $d$ between the $X$ and $Y$ layers, and $Q=(4\pi/\lambda)\sin\theta$ is the modulus of the scattering vector perpendicular to film thickness for an incidence angle $\theta$. In general, reflection coefficients are different when the X-ray impinges from the top, coefficients $r_{X}$, $r_{Y}$, and $r_{YX}$, or from the bottom, coefficients $\bar{r}_{X}$, $\bar{r}_{Y}$, and $\bar{r}_{YX}$. 

Eqs.~(\ref{eq:receqs}) are used recursively, starting from the atomic monolayers within the building blocks $A$ and $B$, whose interlayer distances are given in Table~\ref{tab:1}. For an atomic monolayer, $r_X = {\bar r}_X=-i\Gamma\sum_a\eta_a f_a(Q,E)$  and $t_X=1+i\Gamma\sum_a\eta_a f_a(0,E)$ where $\eta_a$ is the area density of atoms $a$ in the monolayer plane and $f_a(Q,E)=f_a^0(Q)+f_a^{\prime}(E)+if_a^{\prime\prime}(E)$ are their atomic scattering factors with resonant amplitudes for X-ray photons of energy $E$, see section S3 in the Support Information. The parameter $\Gamma=r_e\lambda C/\sin\theta$ arises from the scattering and photoelectric absorption cross sections, and it is very small due to the value of electron radius $r_e = 2.818\times10^{-5}$\,\AA. The $\sin\theta$ takes into account area variation of the beam footprint at the sample surface, and the polarization term $C$ is always equal to 1 for $t_X$, as well as in $r_X$ when using linearly polarized X-rays (most synchrotron facilities) \cite{sm02a}. For accurate curve fitting purposes with unpolarized X-rays, take $C^2= \frac{1}{2}(1+\cos^2 2\theta)$ in $r_X$. Throughout this work, $C=1$ is considered for sake of simplicity.   

\begin{table*}
\centering
\caption{Atomic monolayers (MLs) distances $d$ in structure models of MBT films on BaF$_2$ (111) substrate. Building block $A$ $=$ Te(1)-Bi(1)-Te(2)-Mn(1)-Te(3)-Bi(2)-Te(4) and $B$ $=$ Te(1)-Bi(1)-Te(2)-Bi(2)-Te(3). Subtrate MLs along [111] direction: F(1)-Ba(1)-F(2), distances $d_{\rm BaF}=0.08949$\,nm and $d_{\rm FF}=0.17898$\,nm. In-plane lattice parameters: $a_A = 0.4334$\,nm,  $a_B = 0.4386$\,nm, and $a_S = 0.4384$\,nm. All $d$-values are from bulk \cite{dl13,sn63}. }
\label{tab:1}
\begin{tabular}{cccc}
\hline\noalign{\smallskip}
\multicolumn{2}{c}{building block $A$ (MnBi$_2$Te$_4$)} & 
\multicolumn{2}{c}{building block $B$ (Bi$_2$Te$_3$)} \\
\noalign{\smallskip}\hline\noalign{\smallskip}
MLs & $d$ (nm) & MLs & $d$ (nm)  \\
\noalign{\smallskip}\hline\noalign{\smallskip}
Te(1)-Bi(1) and Bi(2)-Te(4) & 0.17073 & Te(1)-Bi(1) and Bi(2)-Te(3) & 0.17434 \\
Bi(1)-Te(2) and Te(3)-Bi(2) & 0.21532 & Bi(1)-Te(2) and Te(2)-Bi(2) & 0.20331 \\
Te(2)-Mn(1) and Mn(1)-Te(3) & 0.15928 & Te(3):Te(1) ($B\!:\!B$ vdW gap) & 0.26126 \\
Te(4):Te(1) ($A\!:\!A$ vdW gap) & 0.27301 & Te(1)-Te(1) $\rightarrow\, d_{QL}$ & 1.01656  \\
Te(1)-Te(1) $\rightarrow\, d_{SL}$ & 1.36367 & --- & --- \\
\noalign{\smallskip}\hline
\end{tabular}
\end{table*}

After calculating the reflection coefficients $r_A$ and $r_B$ of the $A$ and $B$ building blocks, the reflection coefficients of any sequence of blocks such as $A\!:\!A\!:\!B\!:\!A\!:\!A\!:\!B$ follows straightforward from eqs.~(\ref{eq:receqs}), that is 
\begin{eqnarray}\label{eq:rAAB}
(X\!=\!A,Y\!=\!A) & \rightarrow & r_{AA}\,, \nonumber \\
(X\!=\!B,Y\!=\!AA) & \rightarrow & r_{AAB}\,,\quad{\rm and} \nonumber \\
(X\!=\!AAB,Y\!=\!AAB) & \rightarrow & r_{AABAAB}\,.  
\end{eqnarray}
The $A\!:\!A$ and $B\!:\!B$ vdW gap distances are given in Table~\ref{tab:1}, and the mean value of 0.26713\,nm has been used for the $A\!:\!B$ vdW gap. In most cases, it is necessary to consider the presence of the perfect and thick substrate lattice underneath the film. As detailed elsewhere \cite{sm17}, eqs.~(\ref{eq:receqs}) also provides the dynamical diffraction solution in specular reflection geometry where refraction, rescattering, and photoelectric absorption are taken into account---the impact of rescattering on diffracted intensities can be figured out by suppressing the term $\bar{r}_X r_Y\,e^{2i \varphi}$ in the denominator of the coefficients in eqs.~(\ref{eq:receqs}). The reflection coefficient $r_S$, obtained from eqs.~(\ref{eq:receqs}), for a thick substrate can be included when simulating the X-ray reflectivity curve $R(\theta)$ for a given model by repeating one more step of the recursive procedure, that is $(X\!=\!AABAAB,Y\!=\!S) \rightarrow r_{SAABAAB}$. It provides $R(\theta) = |r_{SAABAAB}|^2$ as the reflectivity curve of the particular example of model in eq.~(\ref{eq:rAAB}) grown on top of a single crystal substrate.

\begin{figure*}
    \centering
  \includegraphics[width=\textwidth]{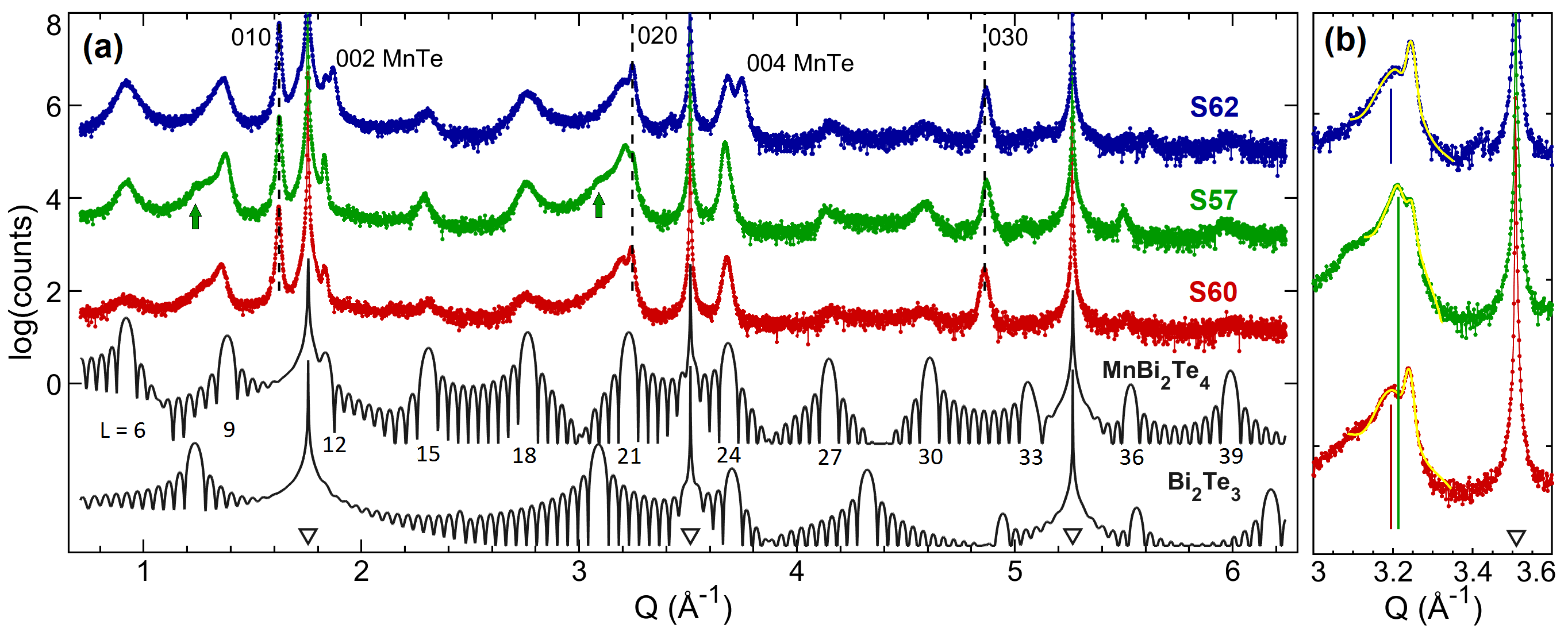}
\caption{(a) $Q$-scans along the surface normal direction in epitaxial MBT films on BaF$_2$ (111) substrate. $Q = (4\pi/\lambda)\sin\theta$. Sample labels (see \S~Materials~and~Methods) are indicated aside each experimental scan. Sharp-high intensity peaks (down triangles) stand for 111, 222, and 333 reflections of the substrate lattice. The 010, 020, and 030 reflections (dashed lines) from a hexagonal Te phase \cite{nb03} are originated by the Te protective capping layer, while the 002 and 004 reflection indexes refer to a hexagonal MnTe phase \cite{hf78} observed in the sample S62 only. Broad peaks are $00L$ reflections ($L=6,\,9,\ldots\,39$) from MBT films. Contributions from Bi$_2$Te$_3$ layers are pointed out by arrows. Simulated curves for MBT and Bi$_2$Te$_3$ single phase 13\,nm thick films are shown at the bottom (out of vertical scale) as reference for peak positions. (b) Zooming of experimental profiles around film $L21$ peak and substrate 222 reflection. $Q$ positions (vertical lines) of the $L21$ peaks obtained by line profile fitting (yellow lines) are indicated: $Q_{21}^{\rm S60} = 3.201\,\textrm{\AA}^{-1}$, $Q_{21}^{\rm S57} = 3.214\,\textrm{\AA}^{-1}$, and $Q_{21}^{\rm S62} = 3.208\,\textrm{\AA}^{-1}$ for samples S60, S57, and S62, respectively.}
\label{fig:xrd}
\end{figure*}

\section{Materials and Methods}
\label{sec:MM}

Thin films of (MnBi$_2$Te$_4$)$_n$(Bi$_2$Te$_3$)$_m$ were grown on freshly cleaved BaF$_2$ (111) substrates using effusion cells charged with Bi$_2$Te$_3$ and phase pure MnTe as prepared by inorganic solid-state reactions \cite{qy19}. The beam equivalent pressure (BEP) is monitored by an ion gauge before and after each growth. The manganese supply is defined by the ratio of the BEPs as $\Phi_{\rm Mn} =  {\rm BEP}_{\rm MnTe}/{\rm BEP}_{{\rm Bi}_2{\rm Te}_3}$. During deposition, the background pressure stays below $5\times10^{-8}$\,mbar, against $7\times10^{-11}$\,mbar base pressure of the growth chamber. BaF$_2$ substrates
were pre-heated at 350$^\circ$C for 10\,min before starting deposition, and kept at 280$^\circ$C during deposition. All samples were prepared with a fixed BEP$_{{\rm Bi}_2{\rm Te}_3}$, resulting in a deposition rate of 0.02\,\AA/s for the
Bi$_2$Te$_3$ cell. The Mn supply controlled through the MnTe effusion cell, providing $\Phi_{\rm Mn} = 0.06$, $0.07$, and $0.11$ for samples labelled S60, S57, and S62, respectively. These samples were covered with a 80\,nm thick Te capping layer to avoid surface oxidation. Another set of samples, without the Te cap, was prepared within similar conditions: samples S27, S34, and S29 with $\Phi_{\rm Mn} = 0.075$, $0.086$, and $0.102$, respectively. For sample S34, the substrate temperature was kept at 300$^\circ$C during deposition. Nominal thickness of the films is close to 20\,nm, except in sample S57 where it is closer to 40\,nm.

XRD measurements were performed with a Bruker high resolution X-ray diffractometer equipped with G{\"{o}}bel mirror, Ge (220) monochromator, and Cu$K_{\alpha1}$ radiation ($\lambda =  1.540562$\,\AA). Transmission electron microscopy (TEM) samples were prepared at the Wilhelm Conrad R{\"{o}}ntgen Research Center for Complex Material System (RCCM) by using Ga$^+$ ion beam milling. Imaging was performed using an uncorrected FEI Titan 80–300 TEM.

\section{Results and Discussions}

Experimental and simulated long-range $Q$-scans along the surface normal direction are shown in Figure~\ref{fig:xrd}. Diffraction peaks of epitaxial MBT films are clearly identified by comparing with the simulated ones. Traces of Bi$_2$Te$_3$ layers are seen in sample S57 (arrows) near peaks $L9$ and $L21$ of the MBT film. Besides film and substrate reflections, there are also diffraction peaks of 010, 020, and 030 reflections from the protective Te capping layer, and two diffraction peaks from a MnTe phase in the S62 sample film grown under higher Mn supply. The presence of MnTe layers epitaxially oriented
with the substrate lattice have already been observed, as well as the formation of multiple MnTe layers inside the \ce{MnBi2Te4} blocks, given rise to blocks composed of 9, 11 or 13 layers \cite{th20}. The presence of MnTe clusters perturb locally the magnetic order of the sample and must be avoided.

\begin{figure*}
    \centering
  \includegraphics[width=.4\textwidth]{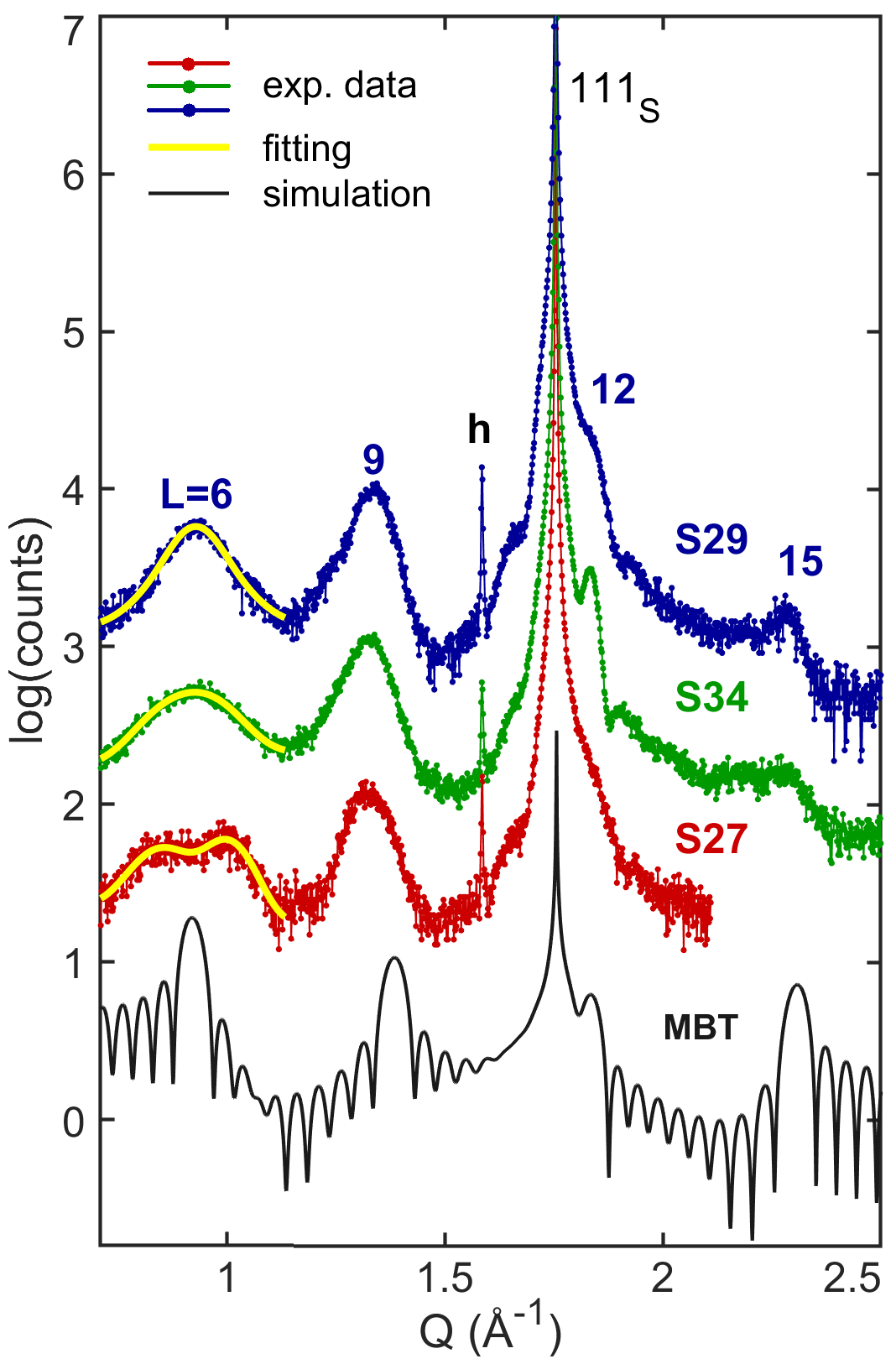}
\caption{$Q$-scans along the surface normal direction in epitaxial MBT films on BaF$_2$ (111) substrate. $Q = (4\pi/\lambda)\sin\theta$. Sample labels (see \S~Materials~and~Methods) are indicated aside each experimental scan (lines with dots). Film $00L$ reflections with $L=6$, 9, 12, and 15 are visible, as well as the 111 substrate reflection. Simulated curve for a 13\,nm thick MBT film is shown at the bottom (out of vertical scale) as reference for peak positions. A hybrid reflection is seen on all scans (letter h); it is a substrate/film reescattering phenomenon extensively discussed elsewhere \cite{sm18}. Line profile fitting (yellow lines) by two gaussians of the diffraction peak L6 provide splitting values of $\Delta Q = 0.15\,\textrm{\AA}^{-1}$, $0.10\,\textrm{\AA}^{-1}$, and null for samples S27, S34, and S29, respectively.}
\label{fig:xrd2}
\end{figure*}

\begin{figure*}
    \centering
  \includegraphics[width=.75\textwidth]{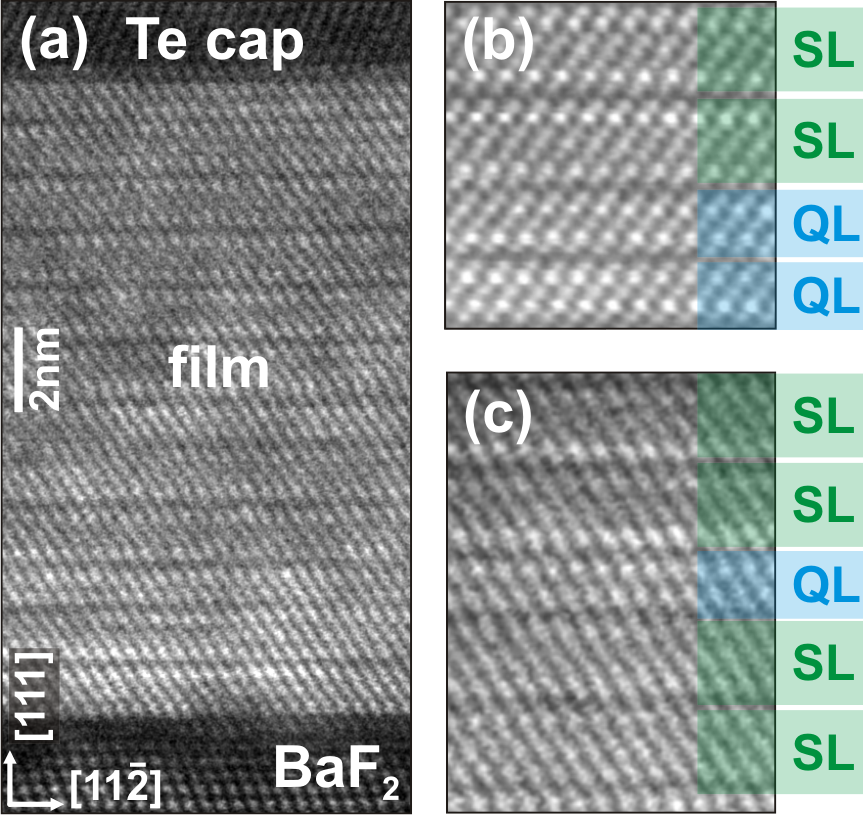}
\caption{Scanning transmission electron microscopy cross-sectional images of a 15 nm thick MBT film capped with Te. (a) Cross-sectional overview image showing the \ce{BaF2} substrate, the epitaxic film, and the Te protective capping layer. (b,c) Detailed views of the film structure showing the coexistence of septuple layers (SLs) and quintuple layers (QLs).}
\label{fig:TEM}      
\end{figure*}

Figure~\ref{fig:xrd2} shows $Q$-scans for the samples without the Te capping layer. No signal from the 002 MnTe reflection can be identified at the right side of the MBT $L12$ peak, even in the case of sample S29 with Mn supply $\Phi_{\rm Mn}=0.102$ that is close to $\Phi_{\rm Mn}=0.11$ used for preparing sample S62 (Figure~\ref{fig:xrd}). However, there are other features evidenced in these intensity curves. As the Mn supply increases, peak $L6$ becomes narrower and peak $L9$ moves slightly towards the expected position of the pure phase. By fitting the line profile of peak $L6$ with two gaussians, they appears set apart by $\Delta Q = 0.15\,\textrm{\AA}^{-1}$ and $0.10\,\textrm{\AA}^{-1}$ in the samples for which $\Phi_{\rm Mn} = 0.075$ and $0.086$, respectively. In the sample for which $\Phi_{\rm Mn}=0.102$, no separation of the gaussians is detected. To clearly understand such features, XRD simulation becomes crucial. Structure models of MBT films based on two building blocks, as depicted in Figure~\ref{fig:ABblocks}d, are supported by TEM images as the one in Figure~\ref{fig:TEM} where only MnBi$_2$Te$_4$ septuple layers (SLs) and Bi$_2$Te$_3$ quintuple layers (QLs) have been observed \cite{pk20}. Bismuth bilayers owing to the deficit of tellurium have been suppressed, probably due to the extra amount of Te from the MnTe source.

\begin{figure*}
    \centering
  \includegraphics[width=\textwidth]{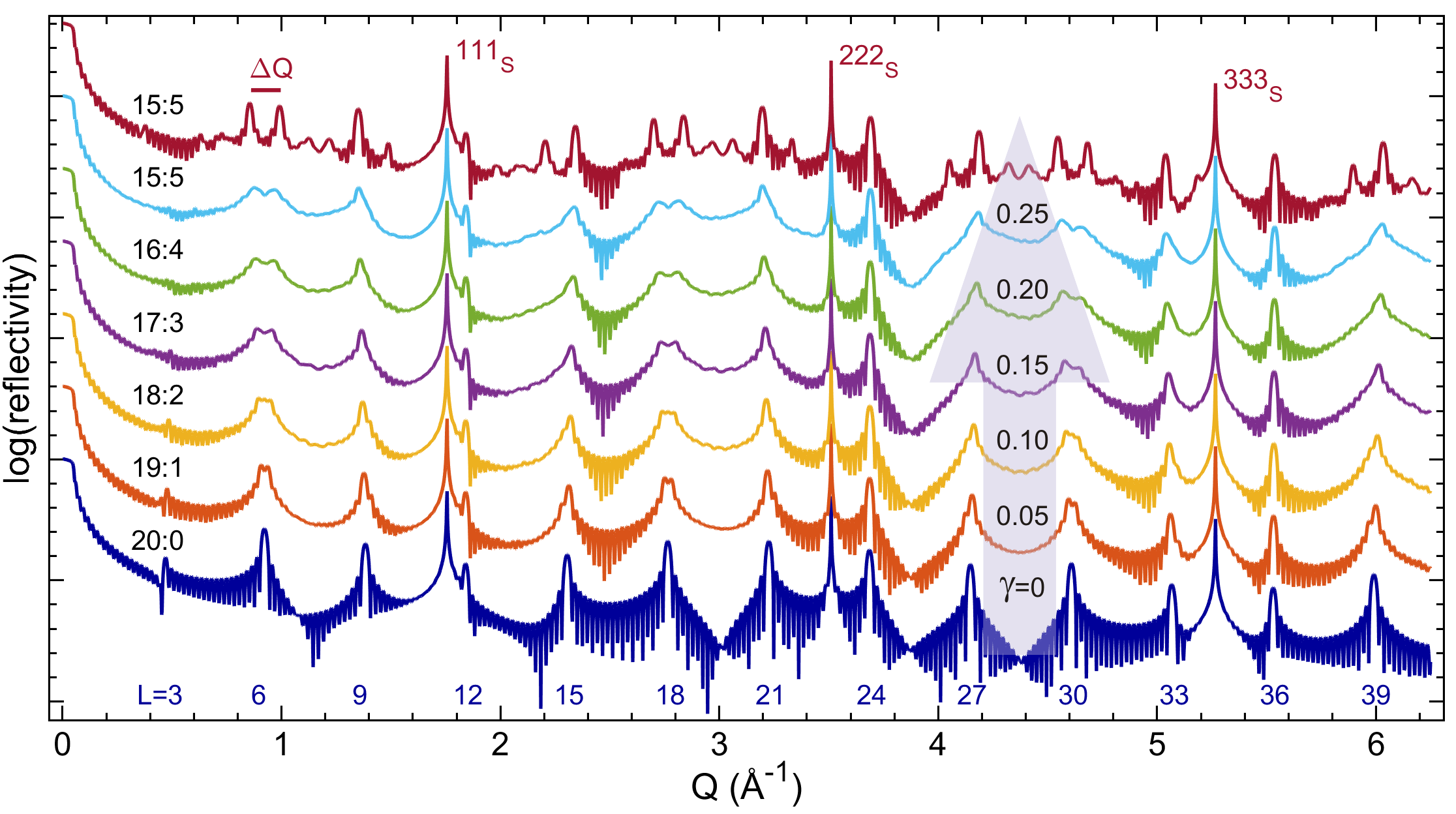}
\caption{Simulated $Q$-scans in epitaxial $({\rm Mn}{\rm Bi}_2{\rm Te}_4)_n({\rm Bi}_2{\rm Te}_3)_m$ films on BaF$_2$ (111) substrate as a function of compositon $n\!:\!m$. Structure models of disordered building blocks according to eqs.~(\ref{eq:PofN}) and (\ref{eq:Gz}) with $\sigma_G=10$ ($\sigma_G=0.1$ for the top curve only). $L$ index of $00L$ reflections in single phase epitaxial MBT films are indicated at the bottom, and substrate reflection at the top. $\Delta Q$ splitting of diffraction peaks $L6$ and $L18$ is proportional to the Mn deficit $\gamma = m/(n+m)$.}
\label{fig:Qscans}      
\end{figure*}

\begin{figure*}
    \centering
  \includegraphics[width=\textwidth]{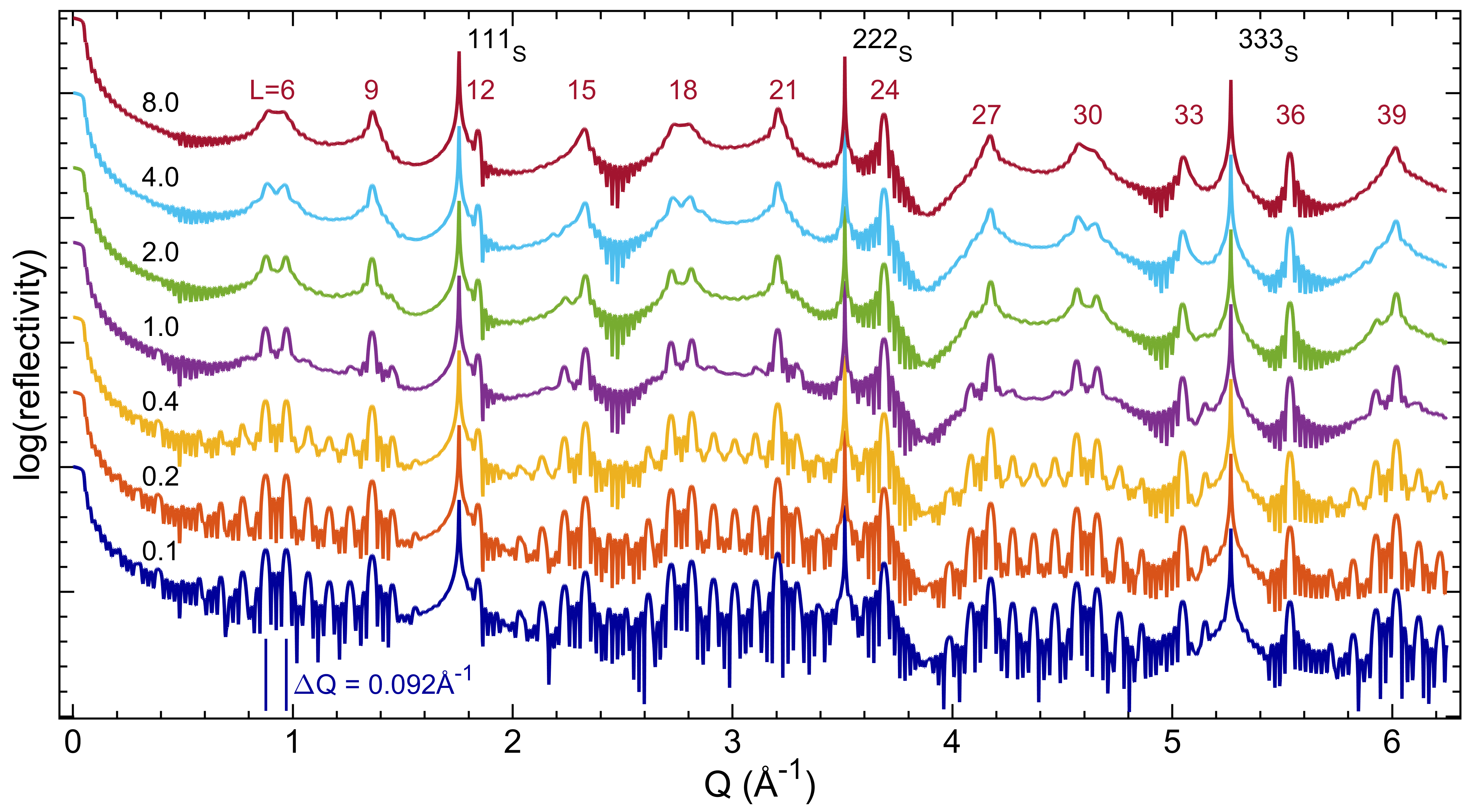}
\caption{Simulated $Q$-scans in epitaxial $({\rm Mn}{\rm Bi}_2{\rm Te}_4)_{16}({\rm Bi}_2{\rm Te}_3)_4$ films on BaF$_2$ (111) substrate as a function of $\sigma_G = 0.1,\,0.2,\ldots,\,8$ in eq.~(\ref{eq:Gz}). Each simulation corresponds to the average curve calculated over an ensemble of 100 models of disordered heterostructures, such as those in Figure~\ref{fig:filmstructure}.}
\label{fig:QscansA16B4}      
\end{figure*}

\begin{figure*}
    \centering
  \includegraphics[width=.75\textwidth]{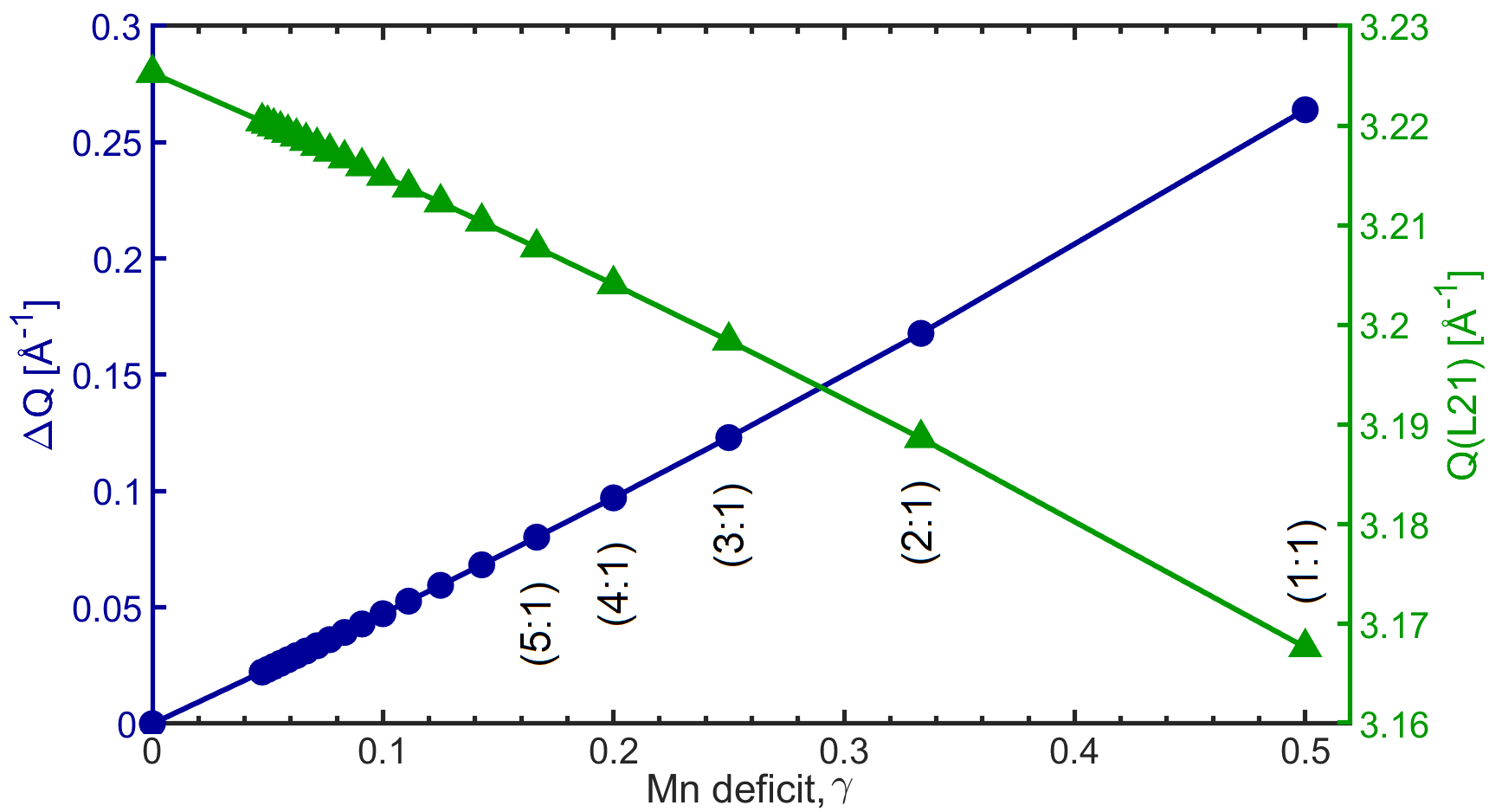}
\caption{Structural parameters accessible from X-ray reflectivity curves as a function of the Mn deficit $\gamma$ in MBT films. $\Delta Q$ (blue solid line and circles, left axis) stands for the separation between adjacent superlattice peaks, eq.~(\ref{eq:DQ}), that are well visible as a splitting of the $L6$ and $L18$ peaks. $Q(L21)$ (green solid line and triangles, right axis) is the $L21$ peak position given by eq.~(\ref{eq:Q21}). $\gamma = m/(n+m)$ for a film with composition $n\!:\!m$ as indicated near a few points.}
\label{fig:DQL21}      
\end{figure*}

Simulated $Q$-scans in disordered $({\rm Mn}{\rm Bi}_2{\rm Te}_4)_{n}$ $({\rm Bi}_2{\rm Te}_3)_m$ films as a function of the Mn deficit $\gamma = m/(n+m)$ are shown in Figure~\ref{fig:Qscans}; low $Q$ regions are detailed in section S1 of the Support Information. For a given degree of disorder, the simulated curves reveal that well visible diffraction peaks as $L6$ and $L18$ undergo a splitting into two superlattice satellite peaks set aparted by a $\Delta Q$ value nearly proportional to the content of QLs (Bi$_2$Te$_3$ blocks) or Mn deficit. Most satellite peaks vanish as the disorder parameter $\sigma_G$ increases, as better seen in Figure~\ref{fig:QscansA16B4}. But, the splitting of the $L6$ and $L18$ peaks remain measurable even in films where the QLs are distributed with high degree of disorder. Within the approximation of stable interlayer distances summarized in Table~\ref{tab:1}, the films have thickness $T = n\, d_{SL} + m\, d_{QL}$, and mean superlattice period $\langle D \rangle = T/m$ where $d_{SL} = 1.36367$\,nm and $d_{QL} = 1.01656$\,nm (Table~\ref{tab:1}). It provides satellite reflections set aparted by
$$\Delta Q(\gamma) = \frac{2\pi}{\langle D \rangle} = \frac{2 \pi m}{n d_{SL} + m d_{QL}} =$$
\begin{equation}\label{eq:DQ}
    =  \frac{2 \pi \gamma}{d_{SL} - \gamma(d_{SL}-d_{QL})}\approx \frac{2\pi}{d_{SL}} \gamma \,.   
\end{equation}
The Mn deficit $\gamma$ also shifts the $L21$ peak position. In MnBi$_2$Te$_4$ film, the lattice parameter $c=3 d_{SL}$, $Q = 2\pi L /c$, and hence for $L=21$, $Q_{21} = 2\pi/\langle d \rangle_0$ as $\langle d \rangle_0 = d_{SL}/7$. In the case of $({\rm Mn}{\rm Bi}_2{\rm Te}_4)_n({\rm Bi}_2{\rm Te}_3)_m$ films, the mean atomic interlayer distance is $\langle d \rangle = T/(7n + 5m)$ where $7n + 5m$ corresponds to the total number of atomic monolayers stacked along the film thickness. It leads to
$$Q_{21}(\gamma) = \frac{2\pi}{\langle d \rangle} =   \frac{2 \pi (7 - 2\gamma)}{d_{SL} - \gamma(d_{SL}-d_{QL})}=$$
\begin{equation}\label{eq:Q21}
    \approx \frac{2\pi}{\langle d \rangle_0}+\frac{2\pi}{\langle d \rangle_0}\left(\frac{5}{7}-\frac{d_{QL}}{d_{SL}}\right)\gamma\,.
\end{equation}
The linear behaviour of $\Delta Q$ and $Q_{21}$ as function of $\gamma$ are shown in Figure~\ref{fig:DQL21}. For small Mn deficit ($\gamma \lesssim 0.25$), the splitting can be taken as  $\Delta Q \approx 0.46\,\gamma\,[\textrm{\AA}^{-1}]$, while the shift in the $L21$ peak position is $\Delta Q_{21}(\gamma) =  Q_{21}(\gamma)-2\pi/\langle d \rangle_0 \approx -0.10\, \gamma\,[\textrm{\AA}^{-1}]$. 

By using the substrate 222 reflection at $Q = 3.5105(\pm0.0001)\,\textrm{\AA}^{-1}$ as reference, the expected $L21$ peak position is at $Q_{21}(0) = 2\pi/\langle d \rangle_0 = 3.2253\,\textrm{\AA}^{-1}$ for unstrained non-tilted epitaxial films with no Mn deficit. In Figure~\ref{fig:xrd}b, the experimental $L21$ peaks have been observed at sligthly different positions, implying in $\Delta Q_{21}^{\rm S60} = Q_{21}^{\rm S60} - Q_{21}(0) = -0.027\,\textrm{\AA}^{-1}$, $\Delta Q_{21}^{\rm S57} = -0.014\,\textrm{\AA}^{-1}$, and $\Delta Q_{21}^{\rm S62} = -0.020\,\textrm{\AA}^{-1}$. According to eq.~(\ref{eq:Q21}), the analyzed films have compositions $x = 0.73$, $0.86$, and $0.80$, respectively. However, these values are reliable as far as the interlayer distances have no dependence with composition, disorder, and film thickness. In highly disordered heterostrucutres, peak broadening is observed instead of peak splitting into superlattice satellite peaks. The X-ray reflectivity simulations in Figures~\ref{fig:Qscans} and \ref{fig:QscansA16B4} clearly show that peaks susceptible to composition are much broader than other peaks, such as the $L24$ peak that has presented a nearly constant width as a function of composition and heterostructure disorder. Therefore, $L6$ peak significantly broader than $L24$ peak can be taken as another evidence, although qualitative, of Mn deficit and disorder in the films. For the sake of comparison, the $L6$ peaks in the S60, S57, and S62 samples have width (fwhm) around $w_{L6} \approx 0.067\,\textrm{\AA}^{-1}$, against nearly half of this value for the $L24$ peak. 

Invariance of $L24$ peak width with composition and disorder implies that their widths $w_{24}^{S60} = 0.0315\,\textrm{\AA}^{-1}$, $w_{24}^{S57} =  0.0260\,\textrm{\AA}^{-1}$, and $w_{24}^{S62} =  0.0386\,\textrm{\AA}^{-1}$ from the experimental $Q$-scans, Figure~\ref{fig:xrd}a, can be used as a measure of the longitudinal (along film thickness) coherence lengths of $20.0(\pm1.0)$\,nm, $24.2(\pm1.2)$\,nm, and $16.3(\pm0.7)$\,nm, respectively. For samples S60 and S62, these lengths are close to the nominal thickness of 20\,nm. But for sample S57, the coherence length is smaller than the nominal thickness of 40\,nm, probably indicating to structural defects unaccounted for in the X-ray diffraction simulation. Main peaks of the Bi$_2$Te$_3$ phase are clearly observed in the $Q$-scan of sample S57, as pointed out by arrows in Figure~\ref{fig:xrd}a. It means that segregation of phases has occurred in this film, which can in part justify a longitudinal coherence length smaller than the total film thickness and also corroborate to MBT phase with composition closer to $x=1$. 

Although vdW epitaxy can take place on substrates with relatively large lateral lattice mismatch \cite{yg15,tg16,lw17,ag17}, it has been demonstrated in Bi$_2$Te$_3$(001) films on BaF$_2$(111) that even mismatch as small as 0.02\% can drastically impact the lateral lattice coherence length, or the lateral size of crystalline domains \cite{sm19}.  In MnBi$_2$Te$_4$ (001) films on BaF$_2$ (111) the lattice mismatch $|\Delta a/a| =|a_A-a_s|/a_S$ is much bigger, of about $1.1$\%. If the amount $\varepsilon$ of lattice misfit that can be elastically accommodated remains within the same order of magnitude, around $5\times10^{-3}$ as observed in Bi$_2$Te$_3$ films, the lateral coherence length in MnBi$_2$Te$_4$ films is probably smaller than 27\,nm ($\approx \frac{a_s}{\varepsilon + |\Delta a/a|}$) \cite{sm19}. An indirect evidence of shorter in-plane coherence length is the absence of thickness fringes around peak $L24$, as smaller domains can lead to more irregular film surfaces smearing out the fringes. The necessary thickness fluctuation for eliminating fringes around the $L24$ peak is discussed in section S2 of the Supporting Information. 

In the case of samples S27, S34, and S29 with exposed films---no cap layers---, the splitting of peak $L6$ in two superlattice peaks as a function of the Mn supply is better evidenced, Figure~\ref{fig:xrd2}. By using eq.~(\ref{eq:DQ}) and the measured values of $\Delta Q$, the Te deficit of $\gamma = 0.30$ and 0.21 are obtained for samples S27 ($\Phi_{\rm Mn} = 0.075$) and S34 ($\Phi_{\rm Mn} = 0.086$), respectively. Both sets of samples, the capped and not capped ones, are indicating that to prevent the formation of undesired MnTe layers by limiting the Mn supply to about $\Phi_{\rm Mn}=0.1$, films with composition close to Mn$_{0.8}$Bi$_2$Te$_{3.8}$ are obtained. This value of $x\simeq0.8$ also corresponds to a limit value of composition detection capability by X-ray diffraction regarding the actual quality of the epitaxial MBT films.

\section{Conclusion}

X-ray diffraction simulation in (\ce{MnBi2Te4})$_n$ (\ce{Bi2Te3})$_m$ structure models as a function of composition and disorder has pointed out a few features that can be promptly exploited in structural analysis of MBT films obtained by molecular beam epitaxy. There are diffraction peaks that split up into superlattice satellites peaks whose separation is directly proportional to the Mn deficit $\gamma$, and another peak whose position can be used to measure the value of $\gamma$. On top of this, there are peaks with line profiles independent of composition and disorder, and can lead to a measure of the longitudinal coherence length. The models used to demonstrate these features were based on a Gaussian probability distribution of the two building blocks present in the $n\!:\!m$ heterostructures. The experimental results showed that increasing the Mn supply provides just a little improvement in composition, with $x$ from around 0.7 to 0.8. The MBT phase with highest composition, $x=0.86$, was observed in the thicker film where some phase segregation is also observed.

\begin{acknowledgments}
We acknowledge financial support from the DFG through No. SFB1170 "Tocotronics" (Projects A01 and C06), No. SFB1143 "Correlated Magnetism," and the W{\"u}rzburg-Dresden Cluster of Excellence on Complexity and Topology in Quantum Matter ct.qmat (EXC 2147, Project No.390858490) and from the BMBF (Project No. 05K19WW2). R.F.S.P., Y.G.C., and S.L.M. acknowledges financial support from FAPESP (Grant No. 2019/01946-1, 2019/11564-9), CNPq (Grant No. 310432/2020-0), and CAPES (finance code 001).
\end{acknowledgments}


\bibliography{ACS_2021manuscript}

\end{document}